\documentstyle[12pt,fullpage]{article}
\newcommand{\alku}{\begin{equation}}
\newcommand{\loppu}{\end{equation}}
\newcommand{\puoli}{\frac{1}{2}}
\newcommand{\rtw}{\frac{1}{\sqrt{2}}}

\newcommand{\eg}{{\it e.g.,\ }}
\newcommand{\ie}{{\it i.e.,\ }}
\newcommand{\etal}{{\it et al.}}

\newcommand{\al}{\alpha}
\newcommand{\be}{\beta}

\newcommand{\de}{\delta}

\newcommand{\th}{\theta}

\newcommand{\cA}{{\cal A}}
\newcommand{\cP}{{\cal P}}

\newcommand{\bfk}{{\bf k}}

\title{\ \ \ \ \ \ \ \ \ \ \ \ \ \ \ \ \ \ \ \ \ \ \ \ \ \
 \ \ \ \ \ \ \ \ \ \ \ \ \ \ \ \ \ \ {\normalsize CTP-2197}\\
  The Ground State Structure and Modular Transformations
  of Fractional Quantum Hall States on a Torus}

\author{Esko Keski-Vakkuri\thanks{Supported in part
by funds provided by the U.S. Department of Energy (DOE)
under contract $\sharp $DE-AC02-76ER03069.} \\
    {\small Center for Theoretical Physics}\\
    {\small Laboratory for Nuclear Science}\\
    {\small and Department of Physics}\\
    {\small Massachusetts Institute of Technology}\\
    {\small 77 Massachusetts Avenue, Cambridge MA 02139}
        \and
        Xiao-Gang Wen\thanks{Supported by NSF grant DMR-9022933.}\\
    {\small Department of Physics}\\
    {\small Massachusetts Institute of Technology}\\
    {\small 77 Massachusetts Avenue, Cambridge MA 02139}}

\date{March, 1993}

\begin{document}

\maketitle

\begin{abstract}

The structure of ground states of generic FQH states
on a torus is studied by using both effective theory
and electron wave function. The relation between the
effective theory and the wave function becomes transparent
when one considers the ground state structure. We
find that the non-abelian Berry's phases of the abelian
Hall states generated by twisting the mass matrix
are identical to the modular transformation matrix
for the characters
of Gaussian conformal field theory. We also show
that the Haldane-Rezayi spin singlet state
has a ten fold ground
state degeneracy on a torus which indicates
such a state is a non-abelian Hall state.

\end{abstract}

\newpage
{\large {\bf 1. Introduction}}

\bigskip

Recently, fractional quantum Hall (FQH) states
were observed in multi-layer
two dimensional electron systems \cite{01}.
The hierarchical structure of the FQH states
in multi-layer systems appears to be different
from that of the single-layer
systems. This indicates that the FQH states
in multi-layer systems may
contain new topological orders. Using the
Chern-Simons effective theory, it
was shown that the possible topological orders
in the abelian FQH states are
classified by a symmetric integer matrix $K$ \cite{02}.
The hierarchical states in the
single-layer systems realize only a small subset
of the possible topological
orders. The multi-layer systems are a natural
place to study more general
topological orders.

As a definition, a generic (abelian) FQH state
with a topological order labeled
by $K$ is described by the following effective theory
\alku
 {\cal L} =\frac{1}{4\pi} K_{IJ} a_{I\mu}\partial_{\nu}
          a_{J\lambda} \epsilon^{\mu \nu \lambda} \ .
\loppu
It was proposed that the multi-layer FQH state \cite{03}
\alku
  \Psi_K=\prod_{i<j,I,J} (z^{(I)}_{i}-z^{(J)}_{j})^{K_{IJ}}
 e^{-\frac{1}{4} \sum_{I,i} \mid z^{(I)}_i
          \mid^2}
\loppu
is described by the above effective theory.
Thus we say that the multi-layer FQH state (2)
has a topological order labeled by $K$. In Eq. (2),
$z_{Ii}=x_{Ii}+iy_{Ii}$ are the
coordinates of the electrons in the $I^{th}$ layer.

By the statement that Eq. (1) is the low energy
effective theory of the FQH state (2)
we mean the following. There exists an energy
scale $E_0$ such that the effective
theory (1) reproduces all excitations of
the FQH state (2) below that energy
scale. At first sight, this statement
appears to be trivial for the FQH states. This is
because the FQH states are incompressible
and there is no excitation below the energy gap.
When det$K\neq 0$, (1) also has a finite gap.
It seems that (1) describes the low energy
exitations (actually no excitation) in (2)
even when $K$ in (1) and (2) are different.
However this naive picture is incorrect.
When we put the FQH states on a compact
Riemann surface, the FQH states will
have a non-trivial ground state
degeneracy (GSD). What is striking is that the GSD
depends on the topology of the space. In
some sense the GSD of the FQH states has non-trivial
``dynamics''. In order to say that
(1) is the effective theory of the FQH state (2),
we have to show that (1) reproduces
the correct ``dynamics'' for the GSD.

We know an universality class describes a class
of systems which flow to
the same infrared fixed point. For every infrared
fixed point we have a
so called low energy effective theory to describe
the systems at or near the
fixed point.
The characteristic effective theory of the
FQH states, as we will see,
is nothing but the Chern-Simons theory.

The low energy effective theory at the
infrared fixed point is much simpler
than the original high energy theory. The effective
theory of the FQH
states, containing only finite degrees of freedom
at low energies
(a consequence of the gap), is the simplest
field theory. The field theory
with finite number of degrees of freedom is
called topological theory and
has attracted a lot of attention after
Witten's pioneer work \cite{04}.

Another question that we are going to
address is how to measure the
matrix $K$ in a physical
way (say, in numerical calculations
and/or in experiments). $K$ as a parameter in the
effective theory is not directly measurable.
We would like to find, as many as possible,
quantum numbers associated with the degenerate
ground states, so that by measuring these
quantum numbers we can extract information
about $K$ and characterize the topological orders
in the FQH state. The results obtained in
this paper will help us to determine the
topological orders from
numerical calculations.

Quantum Hall state is closely related to
the conformal field theory. The
abelian quantum Hall states
correspond to Gaussian models defined
on a lattice characterized by the same matrix $K$.
Such a connection has appeared in the edge
excitations of the Hall states \cite{05}.
To further confirm the above
connection, we studied the transformation
properties of the ground states of the
Hall system under
modular transformations. We find that the
transformation defined by
the non-abelian Berry's phase
of Hall states reproduces the modular
transformation matrices between
the conformal blocks of the
Gaussian model.

In section 2, we study the ground
states of the effective theory (1) on a
torus. We will show that the GSD is equal
to $|\hbox{det}K|$. In section 3 we
will study the
multi-layer FQH state (2) on a torus
and show that the GSD is also given
by $|\hbox{det}K|$.
We further show that the global pieces
of the ground state wave functions of
the effective theory
and the FQH states are identical. This
indicates that (1) is indeed the effective
theory of the FQH states (2) (in the sense
discussed above). We also studied the generalized
hierarchical states and show in section 4
that (1) can also be the effective theory
of the hierarchical
states. In section 5 we consider also a
special FQH state, the $\nu = \puoli$
Haldane-Rezayi state,
and we calculate its GSD on a torus. We
find that GSD = 10, which cannot be explained
by any simple abelian
quantum Hall state. This strongly suggests
that the Haldane-Rezayi state is a non-abelian
Hall state.
In section 6 we study other quantum numbers
of the ground states, in addition to
the GSD and the filling fraction, that provide
information about $K$. The properties of
the ground states under translation are
studied and new quantum numbers can
indeed provide
additional information about $K$. In the
case of a $2\times 2$ matrix $K$, the new
quantum numbers completely determine
the $K$ matrix. In section 7 we calculate
the non-abelian Berry's phase
of the Hall states and discuss its relation
to the modular transformations of the conformal
blocks. In section 8 we use our
results to study the
topological orders in some FQH states.

\bigskip

{\large {\bf 2. The GSD in the Effective Field Theory}}

\bigskip

In this section we investigate the ground state
structure of an effective field theory suggested
to describe excitations around the hierarchy
FQH states. Our analysis follows the lines of \cite{1}.
The action for the effective field theory is
\alku
  S = \int d^{3}x [ \frac{1}{4\pi } K_{IJ}
  a_{I\mu } \partial _{\nu } a_{J\lambda }
  \epsilon^{\mu \nu \lambda}
      + \frac{1}{4M} g^{\mu \alpha} g^{\nu \beta}
    f_{I\mu \nu} f_{I\alpha \beta} ]\ .
\loppu
There are $\kappa$ $U(1)$-gauge fields
$a_{I\mu },\ I = 1,\ldots , \kappa$ with field strengths
$f_{I\mu \nu} = \partial_{\mu} a_{I\nu} - \partial_{\nu}
a_{I\mu}$ living on a spacetime
${\bf R} \times \Sigma$, ${\bf R}$ is
time, the space $\Sigma$ is a torus and $M$ is a
parameter with a dimension of mass.
The coefficients $K_{IJ}$ form a symmetric
$\kappa \times \kappa$-matrix $K = (K_{IJ})$ and all
its elements are integers.
The metric $g$ is of the form
\alku  (g^{\mu \nu}) = \left( \begin{array}{lr}
g^{00} & 0 \\ 0 & (-g^{ij}) \end{array} \right) \loppu
and it (together with the parameter $M$) gives the
scale of exitations of the gauge fields.
\medskip

We will use Weyl gauge $a_{I0} = 0$. Thus we are left
with only the spatial parts of the gauge fields: $a_{Ii}(x),\ i=1,2$.
On a torus the global and local excitations of the
gauge fields can be separated:
\alku
  a_{Ii}(x) = \frac{\theta_{Ii}(x_0)}{L_i} + \tilde{a}_{Ii}(x) \ ;
\loppu
where $L_1\ (L_2)$ is the length of the torus in
$x_1\ (x_2)$-direction and each $\tilde{a}_{Ii}(x)$
satisfies
\alku
  \int_{\Sigma} d^2x \tilde{a}_{Ii}(x) = 0 \ .
\loppu

The gauge invariant physical observables are
\alku e^{i\oint_{C_j} \vec{a}_I \cdot d\vec{x}} =
e^{i\theta_{Ij}}\ ;\loppu where the contour
integral is taken around
one of the homology cycles $C_j,\ j = 1,2$ around the
torus. In order of this to be consistent, each $\theta_{Ij}$
must be periodic:
$\theta_{Ij} + 2\pi = \theta_{Ij}$.

\medskip

The Weyl gauge condition leads to constraints
\begin{eqnarray}
   0 = \frac{\delta S}{\delta a_{I0}} & = &
\frac{K_{IJ}}{4\pi}
      (\partial_i a_{Jj} - \partial_j a_{Ji} )
\epsilon^{0ij} + \frac{1}{M} g^{ij} \partial_i f_{Ioj}
                                             \nonumber \\
                                      & = &
\frac{K_{IJ}}{4\pi} \epsilon^{0ij} \tilde{f}_{Jij} +
\frac{1}{M} g^{ij} \partial_i \tilde{f}_{I0j}
\end{eqnarray}

Because of the condition (8) the action can now be
factorized into
\begin{eqnarray}
   S & = & \int dt \int_{\Sigma} d^2x
         [ \frac{K_{IJ}}{4\pi} \frac{\theta_{Ii}}{L_i}
\partial_0 \frac{\theta_{Jj}}{L_j} \epsilon^{0ij}
           + \frac{1}{2} m^{ij} \partial_0 \frac{\theta_{Ii}}{L_i}
\partial_0 \frac{\theta_{Ij}}{L_j} ] \nonumber \\
     & + & \int d^3x \tilde{{\cal L}} (\tilde{a}_{I\mu} ,
\partial_{\mu} \tilde{a}_{I\nu}) \nonumber \\
     & = & \int dt [\frac{K_{IJ}}{4\pi} (\theta_{I2}
\dot{\theta}_{J1} - \theta_{I1} \dot{\theta}_{J2} ) +
\frac{1}{2} m^{ij} \dot{\theta}_{Ii}
           \dot{\theta}_{Ij} ] + \tilde{S}_{local} \ ;
\end{eqnarray}
where $\tilde{S}_{local}$ is the action for local excitations
and the mass matrix is given by
the metric and the dimensionful parameter $M$:
$m^{ij} = \frac{1}{M} g^{00} g^{ij}$.
We will neglect the local part and concentrate only in
the term in the brackets, which is the Lagrangian of
the global excitations.
(This kind of ``topological'' Lagrangians have been
studied previously in detail {\it e.g.} in \cite{111}
(in planar geometry).)

\medskip

Let us assume that the mass matrix defined above
has an inverse. Then it is easy to move to the Hamiltonian
picture. After solving
for canonical momenta, Legendre transforming and
quantizing we find
\alku
   H = \frac{1}{2} (m^{-1})_{ij} \Sigma_I
(\frac{\partial}{\partial \theta_{Ii}}  - A^{\theta}_{Ii} )
(\frac{\partial}{\partial \theta_{Ij}} - A^{\theta}_{Ij} ) \ \ ,
\loppu
the Hamiltonian that governs the dynamics of
the wavefunction $\psi (\theta_{Ii} )$ of the global
excitations. Formally it describes a particle
moving on a $2\kappa$-dimensional
torus parametrised by $(\theta_{I1} ,\theta_{I2})$,
in a uniform magnetic field produced by the gauge
potential $A^{\theta}_{Ii}$.
It is convenient to make a change of coordinates
such that the mass matrix becomes diagonal. For this we
introduce a complex number $\tau = \tau_x + i\tau_y$ so
that we can rewrite the mass matrix as
\alku m = m_0 \left( \begin{array}{lr} 1 & \tau_x \\
\tau_x & \tau^2_x + \tau^2_y \end{array} \right) \ \ . \loppu
Next we diagonalize its inverse $(m^{-1})_{ij}$ with a matrix
  \alku S = \frac{1}{2\pi} \left( \begin{array}{lr} 1 &
\tau_x \\ 0 & \tau_y \end{array} \right) \ , \loppu
which we use to define new coordinates $(x_I,y_I)$:
 \alku \left( \begin{array}{c} x_I \\ y_I \end{array} \right) =
S \left( \begin{array}{c} \theta_{I1} \\ \theta_{I2} \end{array} \right)
     = \frac{1}{2\pi} \left( \begin{array}{lr} 1 & \tau_x \\
0 & \tau_y \end{array} \right)
    \left( \begin{array}{c}  \theta_{I1} \\ \theta_{I2} \end{array}
\right) \ \ . \loppu
The periodicity of the torus is now reflected in $(x_I+1,y_I)$,
$(x_I,y_I)$ and $(x_I+\tau_x ,y_I+\tau_y )$ being identical
points.
In the new coordinates the Hamiltonian takes the form
\alku
  H = - \frac{1}{2m_0} \Sigma_I [
(\frac{\partial}{\partial x_I} - i A_{Ix} )^2 +
( \frac{\partial}{\partial y_I} - i A_{Iy} )^2 ] \ ;
\loppu
where, using the Landau gauge, the gauge potentials are
\alku
  (A_{Ix} , A_{Iy} ) = \frac{2\pi}{\tau_y} K_{IJ} (-y_J,0)\ .
\loppu

Now we can proceed to find
the general form of the ground state wave function
of the Hamiltonian (14).
We leave the details in Appendix, where we also
discuss the symmetry
properties of (14) and identify the translation generators.
Here we just state that the general form of the
ground state wavefunction
is
\alku
  \psi = f(\{ z_I\}) e^{-\frac{\pi}{\tau_y} K_{IJ} y_I y_J} \ ,
\loppu
where the function $f(\{ z_I \} )$ is a holomorphic
function of complex variables $z_I = x_I + iy_I$. (In terms of the old
variables $z_I = \frac{1}{2\pi}(\theta_{I1} +
\tau \theta_{I2})$ .) In Landau gauge the
wavefunction (16) is quasiperiodic:

\alku \left\{ \begin{array}{l}
  \psi (x_I + 1) = \psi (x_I) \\
  \psi (x_I+\tau_x ,y_I+\tau_y) = \psi (x_I, y_I)
\exp (-i2\pi K_{IJ} x_J -i\pi \tau_x K_{II})\ .
   \end{array} \right.
\loppu
We use the convention of showing explicitly only the translated arguments
of functions. Notice also that we do not sum over the index $I$.
In order to satisfy these conditions, the
holomorphic part must obey
\alku \left\{ \begin{array}{l}
           f(z_I + 1) = f(z_I) \\
           f(z_I + \tau ) = f(z_I) \exp (-i\pi
\tau K_{II} - i2\pi K_{IJ} z_J)\ .
    \end{array} \right.   \loppu
In the special case that $K$ is diagonal, this
reduces to the previously studied case of \cite{1}.
Now we want to ask what is the most general class of
functions that satisfies the conditions above, and in particular,
how many of them are linearly independent. In other
 words, the problem is to find a basis of the
space $V(K,\tau )$ of entire functions $f$ of
$\kappa$ complex variables that satisfy the
periodicity conditions (18) above.

In ref. \cite{3} p. 122-125, there is a slightly less
general problem for entire functions of several
 complex variables. Following the
arguments there, it is fairly straightforward to
generalize the result for our case. One can prove that the
general form of a function $f(\vec{z}) \in V(K,\tau )$ is
\alku
  f(\vec{z}) = \Sigma_{\vec{n}} \chi (\vec{n})
e^{i\pi (K\vec{n}) \Omega K^{-1} (K\vec{n}) +
i2\pi \vec{n} \cdot \vec{z}}\ ;
\loppu
where the matrix $\Omega = \tau I_{\kappa \times \kappa}$,
and $\chi (\vec{n})$ is constant
for cosets $\vec{\alpha} + KZ^{\kappa}$ corresponding to the
coset lattice $Z^{\kappa}/KZ^{\kappa}$. (The
notation $KZ^{\kappa}$ means the
lattice generated by the column vectors of the matrix $K$,
$\vec{z}$ means the $\kappa$-component vector $(z_I)$.)
For instance, choosing $\chi (\vec{n})$ as the characteristic
functions $\chi_{\vec{\alpha}} (\vec{n})$ of cosets
$\vec{\alpha} + KZ^{\kappa}$ (\ie $\chi_{\vec{\alpha}}
(\vec{n}) = 1$ if there is a vector $\vec{m} \in Z^{\kappa}$
such that
$\vec{n} = \vec{\alpha} + K\vec{m}$, otherwise
$\chi_{\vec{\alpha}} (\vec{n}) = 0$ ) gives
a set of basis vectors of the function space $V(K,\tau)$
\alku f(\vec{z}) = f^{K}_{\vec{\alpha}}
(\vec{z} \mid \tau) =
                   \Theta \left[ \begin{array}{c} K^{-1}
\vec{\alpha} \\ \vec{0} \end{array} \right] (K\vec{z} \mid \tau K)
\loppu
labeled by integer quantum numbers
$\vec{\alpha} = (\alpha_1,\ldots ,\alpha_{\kappa})$
which live in the coset space $Z^{\kappa}/KZ^{\kappa}$.
Thus the number of independent basis vectors
corresponds to the number of elements in the
coset space $Z^{\kappa}/KZ^{\kappa}$.
Since the unit cells of the lattice $KZ^{\kappa}$
 have volume $|\hbox{det}K|$ and the unit cells of
the lattice $Z^{\kappa}$ have volume 1, we conclude
that there are $\bfk \equiv |\hbox{det}K|$ linearly
independent holomorphic functions $f(\vec{z})$ that
satisfy the periodicity conditions (18).
Hence the ground state degeneracy in the effective
field theory is $\bfk$.

\medskip
In general, we could compactify the space
into a Riemann surface $\Sigma_g$ with higher
genus $g$.  Without going into
details, we note that using the canonical one
cycles $A_i,B_i;\ i=1,\ldots ,g$ on $\Sigma_g$ and
the respective closed one forms $w_i,\eta_i$ and
expanding the gauge connections $A^I$ in this
basis, the Lagrangian (1) would factorize into $g$
copies of a system analyzed above. Thus we conclude
that the ground state degeneracy of the theory (1)
on a general Riemann surface $\Sigma_g$ (with the real
line as the time coordinate axis) would be $\bfk^g$.

\bigskip

{\large {\bf 3. The Multi-layer Wave Function}}

\bigskip

We have now seen how the degeneracy of
the ground state can be analyzed
using the effective field theory. However,
if the effective field theory
is to describe multi-layer systems, it is important
to see if we can reproduce the same result analyzing
 directly the wave function
itself. To construct the wavefunction on a torus, we
will follow the lines of
Haldane and Rezayi in \cite{4} where they studied
the Laughlin wave function on a torus.

Let us consider a multi-layer
electron system on a torus parametrized by
$0< \xi <2\pi$ and $0 <\eta < 2\pi$.
First let us consider the one-particle wave function.
The wave function satisfies a quasiperiodic
boundary condition
which can be chosen to be
\alku \begin{array}{l}
\psi(\xi+2\pi,\eta)=\psi(\xi,\eta) \\
\psi(\xi,\eta+2\pi)=e^{-iN_\phi (\xi +\tau_x \eta) -
 i\pi N_{\phi} \tau_x}\psi(\xi,\eta)
\end{array}  \ \ .
\loppu
Let us also assume that the electrons have the
mass matrix given by Eq. (11). The mass matrix can
be diagonalized by choosing
a new coordinate
\alku \left( \begin{array}{c} x \\ y
\end{array} \right) ={1\over 2\pi}
\left( \begin{array}{lr} 1 & \tau_x \\ 0 &
\tau_y \end{array}\right)
\left( \begin{array}{c} \xi \\ \eta \end{array} \right)
\loppu
In terms of the new coordinates, and under gauge choice
\alku
(A_x, A_y)=(y{2\pi N_\phi\over \tau_y},0)
\loppu
the electron in the first Landau
level has the following form of the wave function
\alku
\psi(\xi,\eta)=e^{-{\pi N_\phi \over \tau_y}y^2} F(z)
\loppu
where
\alku
z=x+iy={\xi+\tau\eta\over 2\pi}
\loppu
and $F(z)$ is a holomorphic function of $z$ with no poles.
Here $N_{\phi}$ is the
number of the flux quanta on the torus.
The boundary condition for $\psi$ (Eq. (21)) translate
into the
following boundary condition for $F(z)$:
\alku \begin{array}{l}
F(z+1)=F(z) \\ F(z+\tau)=e^{-i(2z+\tau)\pi N_\phi} F(z) \ .
       \end{array}
\loppu

On a plane, a multilayer FQH state is described by the
following type of many-body wavefunction

\alku \psi_K (z^{(I)}_i) = [ \Pi^{\kappa }_{I=1} \Pi^{N_I}_{i<j}
 (z^{(I)}_i - z^{(I)}_j )^{K_{II}} ] \ [ \Pi^{\kappa}_{I<J} \Pi^{N_I}_{i=1}
  \Pi^{N_J}_{j=1} (z^{(I)}_i - z^{(J)}_j )^{K_{IJ}} ]
   e^{-\frac{1}{4} eB \Sigma_{I,i} \mid z^{(I)}_i \mid^2 } \ \ .
\loppu
The index $I$ labels the $\kappa$ different
two-dimensional layers and $z^{(I)}_i$ are the
coordinates of the $N_I$
electrons in the $I^{th}$ layer. For more general
discussion of the wave function,
iterative methods to construct it, and other properties
of multilayered systems,
see {\it eg.} \cite{5}, \cite{6} and \cite{61}.

Our aim is to study this wave function on a
torus, \ie impose periodic
boundary conditions and see what restrictions we
get to the center
of mass part to be added to the wave function.

Let us first for simplicity restrict ourselves to
study the two-layer case ($\kappa = 2$).
We start with a trial wave function of the general form
\alku
 \psi (z_i,w_i) = F(z_i,w_i) e^{-\frac{\pi}{\tau_y} N_\phi
\Sigma_i y^2_i -\frac{\pi}{\tau_y} N_\phi \Sigma_j v^2_j }
\loppu
for electrons with mass matrix (11) in the Landau gauge.
Here $z_i=x_i+iy_i$ ($w_i=u_i+iv_i$) are the coordinates
of the electrons in the layer $I=1$ ($I=2$)
and $F(z_i,w_i)$ is the holomorphic
part of the wave function. We note that all electrons
described by (28) are in the first Landau level.
Generalizing the boundary conditions (21) to the
many-body wave function, we have
\alku \begin{array}{l} \psi (z_i+1) = \psi (z_i) \\
                    \psi (z_i+\tau ) = \psi (z_i) e^{-2i\pi N_\phi x_i
                                        - i\pi N_{\phi} \tau_x} \\
                    \psi (w_i+1) = \psi (w_i) \\
                    \psi (w_i+\tau ) = \psi (w_i) e^{-2i\pi N_\phi u_i
                                        - i\pi N_{\phi} \tau_x} \ \ ,
   \end{array}  \loppu
in agreement with the conventions used in \cite{4} for the single-layer
case.
{}From these conditions we derive the following periodicity
requirements for the holomorphic part $F(z_i,w_i)$:
\alku \begin{array}{l} F(z_i+1) = F(z_i) \\
                    F(z_i+\tau ) = F(z_i)
e^{-i\pi N_\phi \tau - i2\pi N_\phi z_i} \\
                    F(w_i+1) = F(w_i) \\
                    F(w_i+\tau ) = F(w_i)
e^{-i\pi N_\phi \tau - i2\pi N_\phi w_i} \ \ .
   \end{array} \loppu
As in \cite{4}, to describe the multilayer wave
function on a torus,
we expect that the holomorphic part will separate
into a function of the center-of-mass
coordinates of the electrons and a product of odd Jacobi
theta functions\footnote{We use the notation $\theta (z)$ for
the odd theta function $\theta_{\puoli ,\puoli} (z)$ of \cite{3}.}
for the relative coordinates,
\alku
   F(z_i,w_i) = f_c(Z,W) \Pi^{N_1}_{i<j}
\theta (z_i - z_j)^{m_1} \Pi^{N_2}_{i<j} \theta (w_i - w_j)^{m_2}
   \Pi^{N_1,N_2}_{i,j=1} \theta (z_i - w_j)^n \ \ .
\loppu
Here $Z=\Sigma_i z_i$, $W=\Sigma_j w_j$ are the
center-of-mass coordinates of the electrons in the
different layers
and the exponents $m_1,m_2,n$ are related by the
magnetic flux, $N_\phi = N_1m_1+N_2n = N_2m_2+N_1n$.
The problem is now to see what are the periodicity
requirements for the center-of-mass function and what is the most
general function that satisfies them. In particular,
we want to see what will result as the degeneracy of the wave functions.
Using the properties of the theta functions, we
find from the theta function part
\alku
  \begin{array}{l} F(z_i+1) = f_c(Z+1)[\cdots ]
(-1)^{(N_1-1)m_1 + N_2n} \\
                   F(z_i+\tau ) =
f_c(Z+\tau )[\cdots ] (-1)^{(N_1-1)m_1 + N_2n}
                              e^{-i\pi [(N_1-1)m_1 + N_2n]
\tau -i2\pi [(N_1m_1 + N_2n)z_i - m_1Z -nW]} \\
                   F(w_i+1) = f_c(W+1)[\cdots ]
(-1)^{(N_2-1)m_2 + N_1n} \\
                   F(w_i+\tau ) = f_c(W+\tau )[\cdots ]
(-1)^{(N_2-1)m_2 + N_1n}
                              e^{-i\pi [(N_2-1)m_2 + N_1n]
\tau - i2\pi [(N_2m_2 + N_1n)w_i - nZ - m_2W]}
 \end{array} \loppu
Comparing these with (30) we finally find the periodicity
requirements for the center-of-mass part:
\alku \begin{array}{l}
                   f_c(Z+1,W) = f_c(Z,W) \\
                   f_c(Z+\tau ,W) = f_c(Z,W)
e^{-i\pi m_1 \tau - i2\pi m_1Z - i2\pi nW} \\
                   f_c(Z,W+1) = f_c(Z,W) \\
                   f_c(Z,W+\tau ) = f_c(Z,W)
e^{-i\pi m_2 \tau - i2\pi nZ - i2\pi m_2W}
   \end{array} \loppu
This looks familiar - these are the same conditions as (18)
in section 2 ! So we know that
the most general entire holomorphic functions are the
functions $f_c(Z,W) = f^K_{\vec{\alpha}} (Z,W \mid \tau)$
in the space $V(K,\tau )$ with a matrix
\alku
    K = \left( \begin{array}{cc} m_1 & n \\ n & m_2 \end{array} \right) \
\loppu
and the basis functions were given  earlier in formula (20).
Thus we know that there can be
$|\hbox{det}K| = \mid m_1m_2 - n^2 \mid$ linearly independent choices.
Thus we have arrived in the same degeneracy as we found in
the effective field theory calculation. In addition, the
center-of-mass part of the multilayer wave function has
{\em the same form} as the ground state wave functions of the EFT (1).

\bigskip
Generalizing this result to the $\kappa$-layer case is now
straightforward. We start again with a trial wave function
\alku
  \psi (z^{(I)}_i) = F(z^{(I)}_i)
e^{-\frac{\pi}{\tau_y} N_{\phi} \Sigma_{I,i} (y^{(I)}_i)^2} \ \ .
\loppu
and the with the (quasi)periodicity requirements for it
\alku
    \begin{array}{l} \psi (z^{(I)}_i+1) = \psi (z^{(I)}_i) \\
                     \psi (z^{(I)}_i+\tau ) = \psi (z^{(I)}_i)
e^{-i2\pi N_\phi x^{(I)}_i -i\pi N_{\phi} \tau_x} \ \ .
    \end{array}
\loppu
These yield conditions
\alku
  \begin{array}{l} F(z^{(I)}_i+1) = F(z^{(I)}_i) \\
                   F(z^{(I)}_i+\tau ) = F(z^{(I)}_i)
e^{-i\pi N_{\phi} \tau - i2\pi N_{\phi} z^{(I)}_i}
  \end{array}
\loppu
for the holomorphic part.
On the other hand, replacing the factors $(z^{(I)}_i - z^{(J)}_j)$
in the wave function with odd theta functions we
can write
\alku
   F(z^{(I)}_i) = f_c(Z^{(I)})\
[\Pi^{\kappa}_{I=1} \Pi^{N_I}_{i<j} \theta (z^{(I)}_i - z^{(I)}_j)^{K_{II}}]\
       [\Pi^{\kappa}_{I<J} \Pi^{N_I}_{i=1}
\Pi^{N_J}_{j=1} \theta (z^{(I)}_i - z^{(J)}_j)^{K_{IJ}}]\ \ .
\loppu
For this the effects of translations are
\alku
   \begin{array}{l} F(z^{(I)}_i+1) = f_c(Z^{(I)}+1)
[\cdots ][\cdots ] (-1)^{K_{IJ}N_J-K_{II}} \\
                    F(z^{(I)}_i+\tau ) = f_c(Z^{(I)}+\tau )
[\cdots ][\cdots ] (-1)^{K_{IJ}N_J-K_{II}}
         e^{-i\pi (K_{IJ}N_J-K_{II})\tau -i2\pi
(K_{IJ}N_Jz^{(I)}_i-K_{IJ}Z^{(J)})}\ \ .
   \end{array}
\loppu
There are now constraints on $N_I$, $N_\phi = K_{IJ}N_J$
for all $I=1,\ldots ,\kappa$ since
the fluxes through all the layers are of equal size. Inserting
these to the conditions (39) above and
comparing with (37) we find the quasiperiodicity conditions
\alku
    \begin{array}{l} f_c (Z^{(I)}+1) = f_c (Z^{(I)}) \\
                     f_c (Z^{(I)}+\tau ) = f_c (Z^{(I)})
e^{-i\pi K_{II} \tau - i2\pi K_{IJ} Z^{(J)}}
    \end{array}
\loppu
for the center-of-mass part of the wave function. These are again
the same conditions
as (18) in section 2.  Therefore the wave functions $\psi$ are
classified by the linearly independent
functions $f_c=f^K_{\vec{\alpha}}$ in the function space $V(K,\tau )$,
and the degeneracy of the ground state wave function is $\bfk$,
the same result as we
found in section 2.

\bigskip

{\large {\bf 4. The Hierarchy FQH Wave function}}

\bigskip

There are now various different proposals \cite{7},\cite{8} on
the market for the ground state
wavefunctions in the hierarchy scheme \cite{81} of the FQHE. In
this section we will study one of these
different wave functions - Read's proposal in \cite{8} - on a torus.

\medskip
In \cite{8} the hierarchy electron wave
function is written in the form
\alku
  \psi (z^{(0)}_i) = \int \Pi^{\kappa - 1}_{I=1}
\Pi^{N_I}_{i=1} d^2z^{(I)}_i\ \Pi^{\kappa - 1}_{I=0} \Pi_{i<j}
                 (z^{(I)}_i - z^{(I)}_j)^{a_I} \Pi^{\kappa - 1}_{I=0}
\Pi_{i,j} (z^{(I+1)}_i - z^{(I)}_j)^{b_{II+1}}
                 e^{-\frac{1}{4}eB \Sigma_i \mid z^0_i \mid^2} \ \ .
\loppu
Here the $z^{(0)}_i$ are the positions of $N_0$ electrons
and the integrals are over coordinates of quasiparticles
at levels $I=1,\ldots ,\kappa -1$ in the hierarchy, each
level $I$ has $N_I$ quasiparticles at positions
$z^{(I)}_i$. The exponents are $a_0$ (odd,$>0$) ,
$a_I$ (even,$>0$) , $b_{II+1}=\pm 1$ and $b_{\kappa -1 \kappa}=0$.

\medskip
First of all, if we ignore the integrations and look at the
function in the integrand we notice that
it can be written in the form (up to an irrelevant overall sign)
\alku
  \psi (z^{(I)}_i) = [\Pi^{\kappa -1}_{I=0} \Pi^{N_I}_{i<j}
(z^{(I)}_i - z^{(I)}_j)^{K_{II}}]\ [\Pi^{\kappa -1}_{I<J} \Pi^{N_I}_{i=1}
                  \Pi^{N_J}_{j=1}
(z^{(I)}_i - z^{(J)}_j)^{K_{IJ}}]\ e^{-\frac{1}{4}
eB \Sigma_i \mid z^{(0)}_i \mid^2}\ \ ,
\loppu
where we have used a matrix of coefficients
\alku (K_{IJ}) = \left( \begin{array}{ccccc}
a_0 & b_{01} & 0 & \cdots & 0 \\
      b_{10} & a_1 & b_{12} & \ & 0 \\
    \vdots & b_{21} & \ddots & \ & \vdots \\
      \ & \ & \ & \ & \ \\
     0 & \cdots & 0 & b_{\kappa -1\kappa -2} & a_{\kappa -1}
             \end{array} \right) \ \ . \loppu
On the other hand, the exponential part can be
rewritten as
\alku
   e^{-\frac{\pi}{\tau_y} N_{\phi} \Sigma_i (y^{(0)}_i)^2}\ \ ,
\loppu
where $y^{(0)}_i$ is the imaginary part of the electron
coordinate $z^{(0)}_i=x^{(0)}_i+iy^{(0)}_i$.
Notice that this looks now formally exactly like the
multilayer wave function as in previous section, except
that only the electron coordinates have an exponential factor.
This is related to the fact that the
total flux $N_\phi$ for a homogenous ground state is given by
\alku (K_{IJ}N_J) = \left( \begin{array}{c} N_\phi \\
0 \\ 0 \\ \vdots \\ 0 \end{array} \right) = (N_\phi \delta_{I,0})\ \ ,
\loppu
$\ie$only the electrons see a non-zero flux.
We can now put the integrand function on a torus as we did
before. We start with an expression
\alku
  \psi (z^{(I)}_i) = F(z^{(I)}_i)
e^{-\frac{\pi}{\tau_y} N_\phi \Sigma_i (y^{(0)}_i)^2}\ \ .
\loppu
For this we require the periodicity properties
\alku
  \begin{array}{l} \psi (z^{(I)}_i+1) = \psi (z^{(I)}_i) \\
                   \psi (z^{(I)}_i+\tau ) =
\left\{ \begin{array}{l} \psi(z^{(0)}_i)e^{-i2\pi N_\phi x^0_i
          -i\pi N_{\phi} \tau_x}
\ \ (I=0) \\
           \psi(z^{(I)}_i)\ \ (I>0)
                                             \end{array} \right.
  \end{array}
\loppu
because of the property (45). Notice that $\psi$ is now actually
periodic in quasiparticle coordinates.
As before, we write
\alku
 F(Z^{(I)}_i) = f_c(Z^{(I)})\
[\Pi^{\kappa -1}_{I=0} \Pi^{N_I}_{i<j}
\theta (z^{(I)}_i - z^{(I)}_j)^{K_{II}}]\
[\Pi^{\kappa -1}_{I<J} \Pi^{N_I}_{i=1} \Pi^{N_J}_{j=1}
\theta (z^{(I)}_i - z^{(J)}_j)^{K_{IJ}}]\ \ .
\loppu
This has the properties
\alku
  \begin{array}{l} F(z^{(I)}_i+1) = f_c(Z^{(I)}+1)
[\cdots ][\cdots ] (-1)^{K_{IJ}N_J-K_{II}}\\
                   F(z^{(I)}_i+\tau ) = f_c(Z^{(I)}+\tau )
[\cdots ][\cdots ] (-1)^{K_{IJ}N_J-K_{II}}
e^{-i\pi (K_{IJ}N_J-K_{II})\tau -i2\pi (K_{IJ}N_Jz^{(I)}_i - K_{IJ}Z^J)}\ \ .
  \end{array}
\loppu
Using (45) and comparing (49) with (47) we find that the
 center-of-mass function has to satisfy the conditions
\alku
  \begin{array}{l} f_c(Z^{(I)}+1) = f_c(Z^{(I)}) \\
                   f_c(Z^{(I)}+\tau ) = f_c(Z^{(I)})
e^{-i\pi K_{II}\tau -i2\pi K_{IJ}Z^{(J)}}\ \ ,
  \end{array}
\loppu
\ie once again we have arrived at the periodicity
conditions (18) of section 2. The integrands
are classified by the $\bfk$ linearly indepent functions
in the space $V(K,\tau)$.
However, to get the {\em electron} wave function, we must
take into account the integrations over
the quasiparticle coordinates. Nevertheless we find it very
plausible that under certain conditions
(\eg the filling factor $\nu < 1$) these functions are also
linearly independent and they
have the
same $\bfk$ fold degeneracy on a torus. Thus we can argue
that the hierarchy FQH states are also described by
the effective field theory (1).

\bigskip

{\large {\bf 5. The Haldane-Rezayi State}}

\bigskip

In order to describe a FQHE plateau at $\nu = \frac{5}{2}$
seen in recent experiments \cite{9}, Haldane and Rezayi \cite{91}
have proposed a
state which is a spin singlet and has $\nu = \frac{1}{2}$
(or $\nu = \frac{5}{2}$ including a completely filled Landau level).
(See also \cite{92}.)
The state is of the form

\alku
\psi_{HR} (z_i,w_i) = \det (\frac{1}{(z_i - w_j)^2})
\Pi^N_{i<j} (z_i - z_j)^2 \Pi^{N}_{i<j} (w_i - w_j)^2
\Pi_{i,j} (z_i - w_j)^2 e^{-\frac{1}{4} eB
(\Sigma_i (\mid z_i \mid^2 + \mid w_i \mid^2 )} \ .
\loppu

Let us put this state on a torus. We proceed as in the
multi-layer case. First we rewrite
\alku
\psi (z_i,w_i) = F(z_i,w_i)
e^{-\frac{\pi}{\tau_y} N_{\phi} (\Sigma_i y^2_i + \Sigma_i v^2_i )}\ ,
\loppu
where $z_i = x_i + iy_i,\ w_i = u_i + iv_i$. Then we
require $\psi$ to be quasiperiodic \ie it
has to satisfy formulas (36).
As in the multi-layer case, the periodicity
requirements for the holomorphic part are
given by (37).
Next we rewrite the holomorphic part using theta
functions and a separate part for the center-of-mass coordinates,
\begin{eqnarray}
    F(z_i,w_i) & = &f_c(Z,W)\ \det (\frac{\theta_{a,b}(z_i - w_j)
\theta_{a',b'} (z_i - w_j)}{\theta^2_{\frac{1}{2} ,
\frac{1}{2}} (z_i - w_j)} ) \nonumber  \\
               &   & \mbox{} \prod^N_{i<j}
\theta^2_{\puoli ,\puoli} (z_i - z_j)
                     \prod^N_{i<j} \theta^2_{\puoli ,\puoli}
(w_i - w_j) \prod_{i,j} \theta^2_{\puoli ,\puoli} (z_i - w_j) \ .
\end{eqnarray}
This satisfies
\alku
\begin{array}{l}
   F(z_i + 1) = f_c(Z+1) [\cdots ] (-1)^{2(a+a')+2(N-1)+2N} \\
   F(z_i + \tau ) = f_c(Z+\tau ) [\cdots ] (-1)^{2(b+b')+2(N-1)+2N}
 e^{-i\pi [2(N-1)+2N] \tau -i2\pi [(2N+2N)z_i -2Z-2W]} \\
   F(w_i + 1) = f_c(W+1) [\cdots ] (-1)^{2(a+a')+2(N-1)+2N} \\
   F(w_i + \tau ) = f_c(W+\tau ) [\cdots ] (-1)^{2(b+b')+2(N-1)+2N}
 e^{-i\pi [2(N-1)+2N] \tau -i2\pi [(2N+2N)w_i -2W-2Z]} \ .
\end{array}
\loppu
Combining (54) and (37) and using $N_{\phi} = 2N + 2N$, we find
then the periodicity requirements for the c.o.m part:
\alku
\begin{array}{l}
  f_c(Z+1,W) = f_c(Z,W) (-1)^{2(a+a')} \\
  f_c(Z+\tau ,W) = f_c(Z,W) (-1)^{2(b+b')}
e^{-i\pi 2\tau - i2\pi (2Z+2W)} \\
  f_c(Z,W+1) = f_c(Z,W) (-1)^{2(a+a')} \\
  f_c(Z,W+\tau ) = f_c(Z,W) (-1)^{2(b+b')}
e^{-i\pi 2\tau - i2\pi (2W+2Z)} \ .
\end{array}
\loppu
If we ignore the factors $(-1)^{2(a+a')},\ (-1)^{2(b+b')}$ for
a moment, we notice that this looks again like the conditions (18)
in section 2. However, we cannot proceed as before
to say that
\alku
 f_c(Z,W) = f_K (Z,W)\ , \ K = \left( \begin{array}{cc} 2 & 2 \\
2 & 2 \end{array} \right) \ .
\loppu
This is because now $K$ is not invertible which is not allowed.
Instead we rewrite
\alku
  f_c(Z,W) = \tilde{f}_c (\tilde{Z} ,\tilde{W} )\ ,
\loppu
where $\tilde{Z} = Z+W\ ,\ \tilde{W} = Z-W$. This function has
to satisfy
\alku
\begin{array}{l}
 \tilde{f}_c (\tilde{Z} +1,\tilde{W} \pm 1) =
\tilde{f}_c (\tilde{Z} ,\tilde{W})(-1)^{2(a+a')} \\
 \tilde{f}_c (\tilde{Z} +\tau ,\tilde{W} \pm \tau ) =
\tilde{f}_c (\tilde{Z} ,\tilde{W})
    e^{-i\pi 2\tau - 2i\pi(2\tilde{Z})}(-1)^{2(b+b')} \ .
\end{array}
\loppu
We find that the most general solution to this can be factorized as
\alku
 \tilde{f}_c (\tilde{Z},\tilde{W}) = g(\tilde{Z}) h(\tilde{W}) \ .
\loppu
The function $h(\tilde{W})$ is periodic under
$\tilde{W} \mapsto \tilde{W} \pm 1,\ \tilde{W}
\mapsto \tilde{W} \pm \tau$.
Liouville's theorem from complex analysis
then tells us that $h$ has to be a constant.
The function $g(\tilde{Z})$ depends on the phases
$(-1)^{2(a+a')},\ (-1)^{2(b+b')}$. If they
 are equal to 1, the function $g(\tilde{Z})$ is (see \cite{3}, p. 124)
\alku
  g(\tilde{Z}) = g^{++}_{\alpha} (\tilde{Z}) \equiv
            \theta \left[ \begin{array}{c} \frac{\alpha}{2} \\
0 \end{array} \right] (2\tilde{Z} \mid 2\tau )\ , \ \alpha = 0,1 \ .
\loppu
Thus we have found that the center-of-mass function depends only on
the {\em combined} c.o.m coordinate
$\tilde{Z} = Z+W$ and that there are two possible linearly
independent choices for it. The other possible values for the phases
$(-1)^{2(a+a')},\ (-1)^{2(b+b')}$ are 1,-1 ; -1, 1 or both
equal to -1, depending on the combination of theta functions
in the determinant part of the wave function. The conditions (58)
are met by modifying the function $g(\tilde{Z})$ to
be
\alku
\begin{array}{l}
  g^{+-}_{\alpha} (\tilde{Z}) \equiv
g^{++}_{\alpha} (\tilde{Z} + \frac{1}{4}) \\
  g^{-+}_{\alpha} (\tilde{Z}) \equiv
e^{i\pi \tilde{Z}} g^{++}_{\alpha} (\tilde{Z} + \frac{\tau}{4}) \\
  g^{--}_{\alpha} (\tilde{Z}) \equiv
e^{i\pi (\tilde{Z}+\frac{1}{4})}
g^{++}_{\alpha} (\tilde{Z} + \frac{1}{4}(\tau + 1)) \ ,
\end{array}
\loppu
depending on the values of the phases. In each chase
$h$ is still a constant and there are two possible center-of-mass
functions.

Next we need to study the degeneracy arising from
rest of the wave function.
In the $\det(\cdots )$ part of the holomorphic part of the
wavefunction we had a product of two theta functions
\alku
      \theta_{a,b} (z_i - w_j) \theta_{a',b'} (z_i - w_j) \ ,
\loppu
so we have freedom to choose different combinations of
theta functions here.
However, there is one subtlety. In principle we could
replace (62) by an arbitrary
linear combination of products (62) that lead to the same
phases $(-1)^{2(a+a')},\ (-1)^{2(b+b')}$ in (54),(55),(58).
Let us list all possible combinations. If both phases are
equal to -1, the only combination with this property
is $(a,b),(a',b')=(0,\puoli),(\puoli,0)$, so there is no problem
with linear combinations. The situation is the same if
the phases are 1,-1 when the
only possibility is $(a,b),(a',b')=(0,0),(0,\puoli)$
or -1,1 when the only possibility is $(0,0),(\puoli,0)$.
But, if the phases both equal to 1, we have
four possibilities: $(a,b),(a',b')=(0,0),(0,0);(0,\puoli),(0,\puoli);
(\puoli,0),(\puoli,0);(\puoli,\puoli),(\puoli,\puoli)$.
However, some of these products are related by formulas
\alku
\begin{array}{l}
 \theta^2_{0,0} (z) = \cos \omega \
\theta^2_{0,\puoli } (z) +  \sin \omega \ \theta^2_{\puoli ,0} (z) \\
 \theta^2_{\puoli ,\puoli} (z) = \sin \omega \
\theta^2_{0,\puoli } (z) - \cos \omega \ \theta^2_{\puoli ,0} (z) \ ,\\
 \cos \omega \equiv \theta^2_{0,\puoli } (0) / \theta^2_{0,0} (0)\ , \
 \sin \omega \equiv \theta^2_{\puoli ,0} (0) / \theta^2_{0,0} (0)\ .
\end{array} \loppu
(see \cite{3} p. 23). We could still have an arbitrary linear
combination of products $(0,\puoli),(0,\puoli);\newline
(\puoli ,0),(\puoli ,0)$ in the determinant:
\alku
  \det (\frac{\ c\theta^2_{0,\puoli }(z_i - w_j) +
d\theta^2_{\puoli ,0} (z_i - w_j)}{\theta^2_{\frac{1}{2} ,
\frac{1}{2}} (z_i - w_j)} ) \ .
\loppu
One may think that this could make the wave
function {\em infinitely} degenerate since $c,d$
could take any real
values\footnote{Notice that there is a constraint.
If one thinks of (c,d) as points in a plane, there
is a cut for the line
$c/ \sin \omega = -d/ \cos \omega = t \in R $. These
values make the determinant a constant because of (63).}.
However, it is possible to prove the interesting
relation\footnote{This relation holds trivially for $N=1$.
One can then generalize it by an induction
argument to $N>1$ making use of (63), the definition of
determinant and elementary row operations.}
\begin{eqnarray}
 & & \det (\frac{\ c\theta^2_{0,\puoli }(z_i - w_j) +
d\theta^2_{\puoli ,0} (z_i - w_j)}{\theta^2_{\frac{1}{2} ,
\frac{1}{2}} (z_i - w_j)} ) \nonumber \\
  &=& r_N(c,d) \det (\frac{\theta^2_{0.\puoli}(z_i - w_j)}
{\theta^2_{\frac{1}{2} ,\frac{1}{2}} (z_i - w_j)} )
  + s_N(c,d) \det (\frac{\theta^2_{\puoli ,0}(z_i - w_j)}
{\theta^2_{\frac{1}{2} ,\frac{1}{2}} (z_i - w_j)} ) \ ,
\end{eqnarray}
where
\alku \begin{array}{l}
  r_N(c,d) = c(c + d\tan \omega )^{N-1} \\
  s_N(c,d) = d(d + c\cot \omega )^{N-1} \ .
 \end{array}
\loppu
The number $N$ is the same that appears in the formulas (51),(53).
This means that every determinant of type (64) can
always be written as a linear combination of two independent factors.
We have then found that there are only two linearly independent
 combinations with phases equal to 1, in addition to the
three combinations with phases equal to 1,-1; -1,1 and -1,-1.
Thus there are five different contributions from the
antisymmetric part of the wavefunction and two from
the center-of-mass part.
Hence the total degeneracy for the Haldane-Rezayi state
on a torus is {\em ten}. This result is in agreement with
numerical simulations
(ref. \cite{10}). Haldane-Rezayi state does not appear to
correspond to any abelian FQH state. For example, an
abelian state described
by
$$ K = \left( \begin{array}{cc} 3 & 1 \\ 1 & -3 \end{array} \right)
$$
gives rise to a ten fold degeneracy, but does not give
filling fraction $\puoli$. (The filling fraction is $\frac{1}{5}$ .)
Thus the Haldane-Rezayi state
is very likely to be a non-abelian state.

\bigskip

{\large {\bf 6. Representations of Translation Generators
and New Quantum Numbers}}

\bigskip

In this section we study the representations of translation
generators in terms of
the degenerate ground states.

The ground state wave functionals from the EFT were found to
be linear combinations of the basis functionals
\alku
   \Psi_{\vec{\alpha }} = f_{\vec{\alpha}} (\vec{z})
e^{-\frac{\pi}{\tau_y} K_{IJ} y_Iy_J}\ ;
\loppu
\alku
   f_{\vec{\alpha}} (\vec{z}) = \Theta
\left[ \begin{array}{c} K^{-1} \vec{\alpha} \\ \vec{0} \end{array} \right]
        (K\vec{z} \mid K \tau)\ .
\loppu
Here $\vec{\alpha}$ labels the cosets $\vec{\alpha} + KZ^{\kappa}$ and
\alku
  \Theta \left[ \begin{array}{c} K^{-1} \vec{\alpha} \\
\vec{0} \end{array} \right]
   (K\vec{z} \mid K \tau) = \sum_{\vec{m}}
\exp \{ i\pi (\vec{m} + K^{-1}\vec{\alpha} ) \tau K (\vec{m} +
K^{-1} \vec{\alpha})
   + i2\pi (\vec{m} + K^{-1} \vec{\alpha} )\cdot K\vec{z} \} \ \ .
\loppu
Let us discuss the effect of magnetic translations combined
with gauge transformations on the degenerate ground states.
The generators of the transformations act on
the ground state wave functions as
\alku \begin{array}{l} t_{I1} \Psi_{\vec{\alpha}} =
e^{i\varphi_{I1}} \Psi_{\vec{\alpha}} (x_J + K^{-1}_{IJ}, y_J) \\
                 t_{I2} \Psi_{\vec{\alpha}} =
e^{i\varphi_{I2}} e^{i\pi \tau_x K^{-1}_{II} +i2\pi x_I}
\Psi_{\vec{\alpha}} (x_J + \tau_x K^{-1}_{IJ},
                   y_J + \tau_y K^{-1}_{IJ} ) \ ,
    \end{array}
\loppu
where $\varphi_{I1}, \varphi_{I2}$ are some arbitrary phases.
We give an example of $t_{Ii}$'s with this property in the Appendix.
By evaluating the contributions from the holomorphic part
and the exponential part separately and separating
out the common factors, we find using the basis (68)
\alku \begin{array}{l}
 t_{I1} \Psi_{\vec{\alpha}} =
e^{i\varphi_{I1} + i2\pi K^{-1}_{IJ} \alpha_J}
\Psi_{\vec{\alpha}} (\vec{z})\\
 t_{I2} \Psi_{\vec{\alpha}} =
e^{i\varphi_{I2}} \Psi_{\vec{\alpha} +
\vec{\delta}_I} (\vec{z})\\
      \end{array}
\loppu
where the vector $\vec{\delta}_I$ means
a vector whose $I^{th}$ component is 1
and others are zero.
It will turn out to be useful to replace
the labels $\vec{\alpha}$
by new labels $\vec{a} \equiv k K^{-1} \vec{\alpha}$.
Using this notation
we can rewrite (61) as
\alku \begin{array}{l}
 t_{I1} \Psi_{\vec{a}} =
e^{i\varphi_{I1} + i2\pi \frac{1}{k} a_i} \Psi_{\vec{a}} (\vec{z}) \\
 t_{I2} \Psi_{\vec{a}} =
e^{i\varphi_{I2}} \Psi_{\vec{a} +
\vec{\Delta}_I} (\vec{z})
   \end{array}
\loppu
where the vector $\vec{\Delta}_I$ means the $I^{th}$ column vector
of the matrix $kK^{-1}$, $k \equiv \det K$.

Let us make two brief comments.
Since it is
arbitrary which direction in the torus we call
the direction for $t_{I1}$ translation and
which for $t_{I2}$ translation, this symmetry
between $t_{I1}, t_{I2}$ must manifest itself somehow. We could have
defined a different basis from (68) by
taking $\chi_{\vec{\alpha}} (\vec{n}) = \
exp (2\pi i \vec{n} \cdot (K^{-1} \vec{\alpha} )$ in (19) instead.
It is then easy to check that in this basis
the $t_{I1}$ translation and the
$t_{I2}$ translation operate in the opposite way
than above. Thus the symmetry between these two directions
manifests itself in the freedom in choosing the basis
for the ground state wavefunctions.

We could also study the effect of the combined
translations $T_1 \equiv \prod t_{I1}$ and
$T_2 \equiv \prod t_{I2}$ (see Appendix). In basis (68) we find
\alku \begin{array}{l}
 T_1 \Psi_{\vec{a}} =
e^{i\varphi_1 +i2\pi \frac{1}{k} \sum_{I} a_I} \Psi_{\vec{a}} \\
 T_2 \Psi_{\vec{a}} =
 e^{i\varphi_2} \Psi_{\vec{a} + \vec{\Delta}} \ ,
   \end{array}
\loppu
where $\vec{\Delta} = \sum_I \vec{\Delta}_I$.
(In the other basis described above
$T_1$ and $T_2$ would again trade meanings.)
We will use these results to try to extract new
information about the matrix $K$ and to find new
measurable quantum numbers that
help us to classify different (abelian) FQH states.

We begin the search for new quantum numbers by
looking at the relative phases of the $T_1$-quantum numbers.
Above we have seen that the $T_1$-quantum number for a
ground state labelled by a vector $\vec{a}$ was
$e^{i\varphi_1 + i2\pi \frac{1}{k} \sum_I a_i}$. Thus
these quantum numbers for different ground states differ only by a
relative phase factor $e^{i2\pi \frac{1}{k} (\sum_I a^,_I - \sum_I a_I)}$.
We will prove that (at least when $K$ is a $2\times 2$-matrix)
this relative phase factor is always a multiple of a
certain factor $e^{i2\pi \frac{1}{k} \phi}$. More importantly,
we will show that the
number $\phi$ can, at least in principle, be measured by
comparing the relative phases of the $T_1$ quantum numbers. We will
then discuss how this new quantum number can be used in
trying to find the $K$ matrix from the observable quantum numbers.

First we need to know how to find all different
labels $\vec{a}$ for the $k = \det K$ different
ground states. We notice that
all labels $\vec{a}$ can be given as linear combinations
\alku
  \vec{a} (c_1,\ldots ,c_{\kappa}) =
c_1\vec{\Delta}_1 + \ldots + c_{\kappa}\vec{\Delta}_{\kappa} \ ,
\loppu
where $\vec{\Delta}_I$ is the $I^{th}$ column
vector of the matrix $kK^{-1}$.  This can be seen in
the following way. Let $\vec{\alpha}$
label the $k$ cosets $\vec{\alpha} + KZ^{\kappa}$.
Then the corresponding labels $\vec{a}$ are
\alku
\vec{a} (\alpha_1,\ldots ,\alpha_{\kappa}) = kK^{-1} \vec{\alpha} =
\alpha_1 \vec{\Delta}_1 + \ldots \alpha_{\kappa} \vec{\Delta}_{\kappa} \ .
\loppu
If we then take $\vec{a} (c_1,\ldots ,c_{\kappa})$
as in (74) and let the $c_I$ run over all integers, we just
get the same labels (75) again.

Next we notice that since all labels $\vec{a}$ are of
the form (74), the $\vec{a}$-dependent part in
the $T_1$-quantum number (73)
is of the form
\alku
  a \equiv \sum_I a_I =
c_1 \phi_1 + \ldots + c_{\kappa} \phi_{\kappa} \ \ \pmod{k} \ ,
\loppu
where $\phi_I = \sum_J (\Delta_I)_J$ (\ie sum of the
entries in the $I^{th}$ column in $kK^{-1}$).
Temporarily, let us not worry about the modding out
by $k$. The equation (76) is a linear Diophantine
equation for $c_1,\ldots ,c_{\kappa}$ with integer coefficients
$\phi_1,\ldots ,\phi_{\kappa}$.  Thus we can use the
following well-known theorem (see eg. \cite{11} p. 30):
there are integer solutions
$c_1,\ldots ,c_{\kappa}$ to (76) if and only if the
greatest common divisor of the $\phi_I$'s,
$\phi \equiv \gcd(\phi_1,\ldots ,\phi_{\kappa})$, is
a divisor of $a$. This means that if the $c_I$ in
Eqn. (76) are taken to be integers, $a$ will always be a multiple
of $\phi$. Conversely, for any multiple
$a = n\phi \ , n = 0,1,2,\ldots $ there is an integer
solution for the $c_I$.
But, it might be that the modding out by $k$ destroys this picture.
However, at least in the
case of $\kappa =2$ it is very easy to prove that also
$k$ is a multiple of $\phi$, so modding out does not
confuse the above result. Thus we know that
there is at least one ground
state $\psi_{\vec{a}}$ with $a = n\phi$ for
every $n=0,1,\ldots ,\frac{k}{\phi}$ . This means
that by studying the relative phases
of the $T_1$-quantum numbers we notice that they
change by factor $e^{i2\pi \frac{1}{k} \phi}$.
If we also know the GSD = $k$ we may then\footnote{We need
this assumption since usually we
may find several candidates for $\phi$ if we try to guess it
{}from phases $e^{i2\pi \frac{1}{k} \phi}$. However, if we also
know what $k$ is, we can choose the right $\phi$ since
we know that it has to be a divisor of $k$.} find $\phi$.

We have now {\em three} quantum numbers that depend on the
matrix $K$: the filling fraction $\nu$,
the ground state degeneracy
$k = \det K$ and the ``relative phase factor'' $\phi$.
There is however one consistency check to be made.
Actually the matrix K is
defined only up to an equivalence transform $K' = W^TKW$,
where $W \in SL(\kappa ,Z)$ and $(1,1,\ldots, 1)W=(1,1,\ldots ,1)$.
The filling factor and the GSD are independent of these
equivalent transformations. We need to ensure that $\phi$ is also
independent. We do this for a $2\times 2$-matrix.

Let us write the $K$ matrix as
\alku K = \left( \begin{array}{cc} A & B \\
B & C \end{array} \right) \ .
\loppu
Solving the conditions for $W$ we find the
equivalent matrices $K'$ parametrized by an integer $n$ :
\alku
K' = \left( \begin{array}{cc} A + n^2 k\nu + 2n(A-B) &
B + n^2 k\nu + n(A-C) \\
B + n^2 k\nu + n(A-C) & C + n^2 k\nu + 2n(B-C) \end{array} \right) \ .
\loppu
For K (77) the quantum number
$\phi = \gcd(\phi_1,\phi_2) = \gcd(C-B,A-B)$.
We can rewrite $A$ and $C$ as
\alku
   A = r_1\phi + B\ ,\ C = r_2\phi + B
\loppu
with some integer factors $r_1,r_2$.
We solve for the sum $r_1 + r_2$ using
\alku
  k\nu = A - 2B + C = (r_1 + r_2)\phi \
\Leftrightarrow \ r_1 + r_2 = \frac{k\nu}{\phi} \ .
\loppu
Since $\frac{k\nu}{\phi}$ is an integer, $\phi$
is a divisor of $k\nu$.
Therefore $\phi$ is invariant under the
equivalence transformations (78):
\alku
\gcd(A-B+nk\nu,C-B-nk\nu)=\gcd(A-B,C-B) \ .
\loppu
The matrix $K$ (77) depends on three parameters.
Can we therefore recover it from the three quantum
numbers $k,\nu$ and $\phi$ ?
To solve $K$ we would need to find $r_1,r_2$ separately.
This we cannot do directly. However we can use
\alku
   k = AC - B^2 \ \Leftrightarrow \ -\phi r_1^2 + (r_1 + r_2) A =
\frac{k}{\phi} \ .
\loppu
Using (80) we can now think of (82) as a
linear Diophantine equation
\alku
  -\phi r_1^2 + \frac{k\nu}{\phi} A = \frac{k}{\phi}
\loppu
with integer coefficients and $r_1^2 ,\ A$ as unknowns.
To be able to describe the system with a
$2\times 2$ matrix $K$ we first need to find integer solutions
$r^2_1,A$ to (83). This is impossible if
$d\equiv \gcd(-\phi, \frac{k\nu}{\phi})$ is not a divisor
of $\frac{k}{\phi}$. If it is, we can use
the well known algorithm for
solving the equation (83) based on the Euclidian
algorithm for finding the greatest common divisor of
$\phi, \frac{k\nu}{\phi}$.
(The algorithm is described {\it e.g.} in \cite{12}.)
When we find a solution $r^2_{10},A_0$ of (83), all
the other solutions are given by
\alku
  r^2_1 = r^2_{10} + t\frac{k\nu}{d\phi} \ ,
\ A = A_0 + t\frac{\phi}{d} \ ;
\loppu
where $t$ is an arbitrary integer parameter.
However, $r_1^2$ has to be a square of an integer.
If there are solutions
like that, then we can find integer
solutions for $A,B,C$ and the system can be
described by a $2\times 2$-matrix $K$.
The parameter $t$ is also related to the fact that $K$ was
defined up to an equivalence transformation. It is easy
to see that if $r^2_{10},A_0$ is a solution
that leads to integers $A,B,C$, by
choosing $t=2n\sqrt{r^2_{10}} + n^2\frac{k\nu}{\phi}$
in (84) yields the equivalent matrices $K'$ as in (78).

We have thus found that the new quantum number $\phi$
in addition to the filling fraction
and the GSD enables us to {\em completely
classify the abelian FQH states} if they can
be described by a $2\times 2$ matrix. In particular,
since we found in sections 2 and 3
that the global piece of the multi-layer
wavefunction is the same as the one found in the EFT,
we can argue that {\em the second level
(abelian) FQH states} (including many double-layer FQH states)
might also be {\em completely classified}
by using the method outlined above.

\bigskip

{\large {\bf 7. Berry's phase}}

\bigskip

One way to obtain more information about $K$
is to measure the non-abelian
Berry's phase associated with the deformation of
the electron mass matrix. Let $H_\tau$ be the
Hamiltonian of the electrons on a torus
with mass
matrix given by Eq. (11). Assume that at proper
filling fraction the electrons
described by $H_\tau$
form a FQH state labeled by $K$.
Then for each $\tau$, $H_\tau$ has
$\bfk \equiv |\hbox{det}K|$ fold degenerate
ground states $|\Phi_n(\tau)\rangle$, $n=1,\ldots,
\bfk$ and $|\Phi_n(\tau)\rangle$ are normalized.
We notice that $H_\tau$ and $H_{\tau+1}$
actually describe the
same system, because $m^{-1}(\tau)$ and
$m^{-1}(\tau+1)$ are related by a
coordinate transformation $(\xi,\eta) \to (\xi-\eta,\eta) $.
Similarly one can show
that $H_\tau$ and $H_{-1/\tau}$
describe the same system due to the
transformation $(\xi,\eta) \to (\eta, -\xi) $.
Therefore
$|\Phi_n(\tau)\rangle$, $|\Phi_n(\tau+1)\rangle$
and $|\Phi_n(-1/\tau)\rangle$ span the same
Hilbert space and
$|\Phi_n(\tau)\rangle$, $|\Phi_n(\tau+1)\rangle $,
and $|\Phi_n(-1/\tau)\rangle$
are related by unitary transformations.

The non-abelian Berry's phase is an unitary matrix
that is associated with an
adiabatic deformation of the Hamiltonian $H_\tau$ \cite{121} .
The deformation starts and
ends with the same Hamiltonian. Let us
denote the deformation path by
$\tau(t)|_{t=0}^1$.
Then the matrix of the non-Abelian Berry's phase is given by
\alku
W[\tau(t)]= \cP {\rm exp}[-i\int_0^1 A(t) dt] W'
\loppu
where $\cP$ denotes the path ordered product, $A$
is a matrix defined by
\alku
A_{nm}(t)=i\langle \Phi_n[\tau(t)]|\;{d\over dt}\;|\Phi_m[\tau(t)]\rangle
\loppu
and $W'$ is the unitary matrix given by
\alku
W'_{nm}=\langle\Phi_n[\tau(1)]|\Phi_m[\tau(0)]\rangle
\loppu

Before going into detailed calculations, let us summarize some
general results.
The non-abelian Berry's phases induced by
the FQH states have the following
special properties. For the path that starts and ends with the
same $\tau$ (\ie
$\tau(0)=\tau(1)$) the non-abelian Berry's
phase (denoted as $W(\tau,\tau)$)
is a pure phase:
\alku
W(\tau,\tau)_{mn}=e^{i\th}\de_{mn}
\loppu
The value of $\th$ may differ from path to path.
If the path connects
$\tau$ and $\tau+2$ (\ie $\tau(1)=\tau(0)+2$) then
the corresponding
non-Abelian Berry's phase (denoted as $W(\tau,\tau+2)$)
is longer a pure phase.
However only the phase of $W(\tau,\tau+2)$
depends on different choices of
paths connecting $\tau$ and $\tau+2$.
Thus $W(\tau,\tau+2)$ can be written
in the following form
\alku
W(\tau,\tau+2)=e^{i\th}U
\loppu
where $U\in SU(\bfk)/Z_{\bfk}$ is
independent of the paths. Similarly the
non-Abelian Berry's phase
$W(\tau, -1/\tau)$ associated with paths
connecting $\tau$ and $-1/\tau$
has a form
\alku
W(\tau,-1/\tau)=e^{i\th}S
\loppu
Again $S\in SU(\bfk)/Z_{\bfk}$ is
independent of the paths.

Because the two matrices $U$ and $S$
are path independent, they reflect the intrinsic
properties of the FQH state. $U$ and $S$,
as $\bfk \times \bfk$ matrices,
contain a lot of information about the
topological orders. In particular
they contain the information about the matrix $K$.

We would like to mention that the two
transformations $\tau\to \tau+2$ and
$\tau\to -{1\over\tau}$ generate a sub-group $\Gamma_2$
of the so called moduli group
\alku
\tau \to {a\tau+b\over c\tau+d}
\loppu
where $a,b,c,d,\in Z$ are integers and $ad-bc=+1$, \ie
\alku
\left(\matrix{ a&b\cr
c&d\cr}\right) \in SL (2,Z).
\loppu
The pair $\{ U,S\}$, being associated with the generators
$\left(\matrix{ 1&2\cr
0&1\cr}\right)$ and $\left(\matrix{0&1\cr
-1&0\cr}\right)$ of $\Gamma_2$, generates
a $\bfk$ dimensional projective representation of $\Gamma_2$.
We see that the topological orders in the FQH states
are closely related to the projective
representations of the moduli group.

First let us prove that the non-abelian
 part of the non-abelian Berry's
phase is path independent. We will use
the multilayer FQH wave function
as an example to perform our calculation.
The wave function in section 3 was constructed
under the gauge $(A_x, A_y)=(y{2\pi N_\phi\over \tau_y},0)$.
To calculate
the Berry's phase it is convenient to
choose a different gauge
\alku
(A_\xi, A_\eta)=(\eta {N_\phi\over 2\pi },0)
\loppu
The new gauge condition is independent of $\tau$.
We would like to remind the reader that the
torus is parametrized by
$0< \xi <2\pi$ and $0 <\eta < 2\pi$. The two
sets of coordinates $(x,y)$ and $(\xi,\eta )$ are related
through $x+iy\equiv z=\xi+\tau\eta$.
Under the new gauge choice
the multilayer wave function has a form
\alku
\psi(z^{(I)}_i)=F(z^{(I)}_i)
e^{{i\tau\over 4\pi}N_\phi \sum_{I,i}(\eta^{(I)}_i)^2}
\loppu
{}From the boundary condition of $\psi$:
\alku
\psi(\xi^{(I)}_i+2\pi\de_{ij},\eta^{(I)}_i)=
\psi(\xi^{(I)}_i,\eta^{(I)}_i),\ \
\psi(\xi^{(I)}_i,\eta^{(I)}_i+2\pi\de_{ij})=
e^{-iN_\phi \xi^{(I)}_j}\psi(\xi^{(I)}_i,\eta^{(I)}_i).
\loppu
one can show that the $F(z^{(I)}_i)$ satisfies exactly
the same boundary condition in section 3 (Eq. (37)).
Thus, repeating the previous calculation,
we find that $F$ is given by (38) with
the center-of-mass wave function given by (20).
We will denote the electron
wave function as $\psi_{\vec{\al}}(\xi^{(I)}_i,\eta^{(I)}_i|\tau)$
if the  center-of-mass wave function
is chosen to be $f_c=f_{\vec{\al}}^K$ in Eq. (20).
We stress that the wave function
$\psi(\xi^{(I)}_i,\eta^{(I)}_i|\tau)$ does
not depend
on $\tau^*$ as one can see from Eq. (38) and (20).
This fact is very important
for the following discussions.

We first notice that the above
degenerate ground state wave
functions $\psi_{\vec{\al}}(\tau)$ (See Eq. (71))
form an represetation of the
generalized magnetic translation
group. This is because the wave functions
of the center-of-mass coordinates
are just those discussed in section 2.
$\psi_{\vec{\al}}(\tau)$ and $\psi_{\vec{\al}'}(\tau)$
are orthogonal to each other because
they carry different quantum numbers of the
commuting unitary magnetic translations $t_{I1}$.
$\psi_{\vec{\al}}(\tau)$ and $\psi_{\vec{\al}'}(\tau)$
have the same
norm since they can be transformed
into each other by the
unitary magnetic translations $t_{I2}$.
Thus we have
\alku
\langle \psi_{\vec{\al}}(\tau)|\psi_{\vec{\al}'}(\tau)\rangle=
g(\tau,\tau^*)\delta_{\vec{\al}\vec{\al}'}
\loppu
{}From (86) and (96) we have
\begin{eqnarray}
(\cA_\tau)_{\vec{\al}\vec{\al}'} & = &
i\int \prod (d\xi_i^{(I)} d\eta_i^{(I)})
{1\over\sqrt{g(\tau,\tau^*)}}
\psi_{\vec{\al}}^*(\tau)
{\partial\over\partial\tau}\left[{1\over \sqrt {g(\tau,\tau^*)}}
\psi_{\vec{\al}'}(\tau) \right] \nonumber \\
& = & i\sqrt{g(\tau,\tau^*)}\ {\partial\over\partial \tau}
 {1\over\sqrt{g(\tau,\tau^*)}} \de_{\vec{\al}\vec{\al}'} +
i {1\over g(\tau,\tau^*) }
 \int d^2\th \psi_{\vec{\al}}^*
{\partial\over\partial \tau} \psi_{\vec{\al}'}.
\end{eqnarray}
Since $\psi_{\vec{\al}}$ is holomorphic in $\tau$,
the above can be rewritten as
\begin{eqnarray}
(\cA_\tau)_{\vec{\al}\vec{\al}'} & = &
i\left[-\puoli {\partial\over\partial\tau}
\ln g(\tau,\tau^*)
+{1\over g(\tau,\tau^*)} {\partial\over\partial\tau}
g(\tau,\tau^*)\right]
\de_{\vec{\al}\vec{\al}'} \nonumber \\
& = & i\de_{\vec{\al}\vec{\al}'} \puoli
{\partial\over\partial\tau} \ln g(\tau,\tau^*).
\end{eqnarray}
Similarly we find that
\alku
(\cA_{\tau^*})_{\vec{\al}\vec{\al}'} =
i\de_{\vec{\al}\vec{\al}'} \left(-\puoli\right)
{\partial\over\partial\tau^*}
\ln g(\tau,\tau^*) .
\loppu
(98) and (99) indicate that the path
ordered product in the definition of the
non-abelian Berry's phase (85) only contribute
to the abelian phase. The
non-abelian part of Berry's phase completely
comes from the relation
between the initial and the final states in Eq. (86).

As we have mentioned, $H_{\tau}$
and $H_{\tau+2}$ describe the same system
after a coordinate transformation and a gauge
transformation. In the fact
if we define a unitary transformation $\hat u$
through (for single-particle
wave function)
\alku
\hat u \psi(\xi^{(I)}, \eta^{(I)})=
e^{{i\over 2\pi} \sum \eta^{(I)} K_{IJ} \eta^{(J)}}
\psi(\xi^{(I)}+2\eta^{(I)},\eta^{(I)})
\loppu
we can show that $H_{\tau+2}=\hat{u} H_\tau \hat{u}^{-1}$. Thus
$g^{-1/2}(\tau+2,\tau^*+2)\psi_{\vec{\al}}(\tau+2)$ and
$g^{-1/2}(\tau,\tau^*)\hat{u} \psi_{\vec{\al}}(\tau)$
are related by a
unitary matrix. The matrix can be
calculated from the transformation properties
the $\Theta$-functions
under the modular transformation $\tau\mapsto \tau +2$.
We find that the non-abelian Berry's phase
associated with
$\tau\mapsto \tau +2$ is given by
\alku
U_{\vec{\al}\vec{\be}}=e^{i\phi_U}
\de_{\vec{\al}\vec{\be}}e^{i2\pi \vec{\al} K^{-1} \vec{\al}}
\loppu
where $\phi_U$ is the path dependent $U(1)$ phase.
Similarly introducing $\hat{s}$
\alku
\hat{s} \psi(\xi^{(I)}, \eta^{(I)})=
e^{-{i\over 2\pi} \sum \xi^{(I)} K_{IJ} \eta^{(J)}}
\psi(\eta^{(I)},-\xi^{(I)})
\loppu
we find that $H_{-1/\tau}=\hat{s} H_\tau \hat{s}^{-1}$.
The unitary matrix
relating
$g^{-1/2}(-1/\tau,-1/\tau^*)\psi_{\vec{\al}}(-1/\tau)$ and
$g^{-1/2}(\tau,\tau^*)\hat{s} \psi_{\vec{\al}}(\tau)$
can be again calculated
{}from the modular transformation $\tau\mapsto -1/\tau$
of the $\Theta$-functions.
The non-abelian Berry's phase
associated with $\tau\mapsto -1/\tau$ is given by
\alku
S_{\vec{\al}\vec{\be}}=e^{i\phi_S}
{1\over \sqrt{\bfk}} e^{i2\pi \vec{\al} K^{-1} \vec{\be}}
\loppu

Up to an overall constant phase, we find
that the eigenvalues of $U$
coincide with statistics of the
quasiparticles the QH state. Thus
the quasiparticle statistics can be
determined through the non-abelian
Berry's phase without even creating a single quasiparticle.

We would like to point out that the above
non-abelian Berry's phase (up to a $U(1)$ phase)
is closely related to the modular
transformations of the Gaussian model in the conformal field
theory. Consider a Gaussian model with $\kappa$ (real)
boson fields $\phi_i$:
\alku
 S = \frac{1}{2\pi} \int d^2z \
\partial_z \phi_i \partial_{\bar{z}} \phi_i \ \ ,
\loppu
where $\phi_i$ parametrize a $\kappa$-dimensional torus,
\ie $\phi_i$ and $\phi_i + 2\pi R_{ij} l_j$ are identified,
$l_j$ are integers and $R_{ij}$ is a real
symmetric matrix. The partition function of the
above Gaussian model is given by
\alku
 Z = \frac{1}{(\eta (\tau) \eta^* (\tau))^{\kappa}}
\sum_{(p,\bar{p})\in \Gamma_R}
             e^{i\pi\sum_{i} (\tau p^2_i - \tau^* \bar{p}^2_i)}
\loppu
where $\Gamma_R$ is the lattice
\alku
 \Gamma_R = \{ (p_i,\bar{p}_i) =
(\puoli (R^{-1})_{ij}m_j + R_{ij}n_j,
\puoli (R^{-1})_{ij}m_j - R_{ij}n_j);\ n_j,m_j \in Z^{\kappa} \} \ .
\loppu
When $(R^{-2})_{ij}$ is an integer matrix $K_{ij}$
with even elements, one can show that the
 partition function can be written
as
\alku
 Z = \frac{1}{(\eta (\tau ) \eta^*(\tau ))^{\kappa}}
\sum_{\vec{\alpha}} \chi^*_{\vec{\alpha}}
(\tau ) \chi_{\vec{\alpha}} (\tau )
\loppu
where $\vec{\alpha} \in Z^{\kappa}/KZ^{\kappa}$
and $\chi_{\vec{\alpha}}$ is given by
\alku
 \chi_{\vec{\alpha}} (\tau ) = \sum_{\vec{n} \in Z^{\kappa}}
e^{i\pi(K\vec{n} + \vec{\alpha}) \tau K^{-1} (K\vec{n} + \vec{\alpha})} \ .
\loppu
The characters of the Gaussian model are given
by $\eta^{-\kappa} (\tau ) \chi_{\vec{\alpha}} (\tau )$.
We see that the number of the
characters (or number of the conformal blocks) is
equal to $\bfk \equiv |\hbox{det}K|$ , the ground state
degeneracy of the Hall state on
a torus.
We also notice that the $\chi_{\vec{\alpha}} (\tau )$
is nothing but the function $f^K_{\vec{\alpha}} (\vec{z} \mid \tau )$
in Eqn. (20)
at $\vec{z} = \vec{0}$. Thus it is not hard to
see that, up to a $U(1)$ phase, the modular transformation of
the characters is given by
the matrices $U$ and $S$ in Eq. (101) and (103)
which are the non-abelian Berry's phases of the Hall state.
This result illustrates the close relation
between the Hall states and the conformal field theory.

\bigskip

{\large {\bf 8. Examples}}

\bigskip

In this section we analyze the new quantum
 numbers for some known FQH states.
As a first example we study the abelian $\nu =\frac{1}{2}$
state described by a matrix
\alku
K = \left( \begin{array}{cc} 3 & 1 \\
1 & 3 \end{array} \right) \ .
\loppu
In this case the GSD is equal to 8. The
vectors $\vec{\alpha}$ labeling the ground states (67)
are $\vec{\alpha} = (0,0);
(1,1);(1,2);(2,1);(2,2);(2,3);(3,2);(3,3)$ and the respective
$\vec{a}$-labels are $\vec{a}=(0,0);(2,2);(1,5);(5,1);(4,4);
(3,7);(7,3);(6,6)$. The quantum number $\phi=2$ and
the values of $a$ as in (76) are indeed its
multiples, $a=0,4,6,6,0,2,2,4$
for the above labels. The action of the translation
generators (72) is shown in Fig. 1a and the
action of the $T_2$ generator
in Fig. 1b. Note that we can block diagonalize
the eight-state representation by defining a
new basis $(0,0)\pm (4,4);
(2,2)\pm (6,6); (5,1)\pm (1,5); (7,3)\pm (3,7)$.
In this basis the representation decomposes into
four blocks of two states (see Fig. 1c).
We can then label these basis states using
the $T_1,T^q_2$ ($q$ comes from filling factor $\nu = \frac{p}{q}$,
here $q=2$) quantum
numbers (since these generators commute). In our
example these quantum numbers form a lattice
of Fig. 1d and they uniquely
label the states.
Moreover, the four blocks in the new basis can
be labelled by their $T^2_1,T^2_2$ quantum numbers.
If we represent this result in a lattice
with $T^2_1,T^2_2$ as $x,y$-axes,
we find that the four blocks sit in a square (Fig. 1e).

As a second example we look at the $\nu =\frac{1}{3}$
state defined by
\alku
K = \left( \begin{array}{cc} 5 & 3 \\ 3 & 3 \end{array} \right) \ .
\loppu
This state has GSD$=6$, the $\phi = 2$, the
labels $\vec{\alpha} = (0,0);(1,1);(2,2);(3,2);(4,3);(5,4)$,\\
$\vec{a} = (0,0);(0,2);(0,4);(3,1);(3,3);(3,5)$ and $a = 0,2,4,4,0,2$.
The representation of (72) is show in Fig. 2a and
the representation of (73) in Fig. 2b. The representation
separates now in two blocks.
In this basis the quantum numbers of $T_1,T^3_2$ form  a
lattice of Fig. 2c.
The $T^3_1,T^3_2$ quantum numbers of the blocks are shown in Fig. 2d.
Again we find that they sit in a square-like pattern.
(It might be that the following is a general feature of the
abelian FQH states: if one block diagonalizes the $T_1,T_2$
representation, the
$T^q_1,T^q_2$ quantum numbers for the blocks
always form a square-like pattern
(with just one point as a special case).
We cannot prove this argument yet, however we found after
studying several different $K$-matrices
that this always seems to be the case.)

\bigskip

{\large {\bf Acknowledgements}}

\bigskip

E. K-V. would like to thank M. Crescimanno
and prof. R. Jackiw for comments in the early stage of this work,
and prof. S.D. Mathur for encouragement.

\bigskip

{\large {\bf Note Added}}

\bigskip

At the completion of this work we received a preprint \cite{13}
with some overlap to our results. We have also been informed by
Dingping Li that hierarchical wave functions on a torus have also
been studied in \cite{14} with overlap to our results.

\bigskip

{\large {\bf Appendix}}

\bigskip

We elaborate on finding the ground state
wavefunctions of the Hamiltonian (15),
\alku
  H = - \frac{1}{2m_0} \Sigma_I [ (\frac{\partial}{\partial x_I} -
i A_{Ix} )^2 + ( \frac{\partial}{\partial y_I} - i A_{Iy} )^2 ] \ ;
\loppu
where the in the Landau gauge the gauge potentials are
\alku
  (A_{Ix} , A_{Iy} ) = \frac{2\pi}{\tau_y} K_{IJ} (-y_J,0)\ .
\loppu
Le us first change to complex coordinates $z_I = x_I + iy_I$.
We define
\alku \left\{ \begin{array}{l}
   \partial_{Iz} = \frac{1}{2} (\partial_{Ix} - i\partial_{Iy} )\ ,
\ A_{Iz} \ =\  \frac{1}{2} (A_{Ix} - iA_{Iy} ) \\
   \partial_{I\bar{z}} = \frac{1}{2} (\partial_{Ix} +
i\partial_{Iy} )\ ,\ A_{I\bar{z}} \ =\  \frac{1}{2} (A_{Ix} + iA_{Iy})\ .
\end{array} \right. \loppu
After some algebra we find $H$ in the form
\alku
  H \ \propto \ \Sigma_I D_I \bar{D}_I \ +\ constant\ ;
\loppu
where
\alku
  D_I = \partial_{Iz} - iA_{Iz} \ \ , \ \ \bar{D}_I =
\partial_{I\bar{z}} - iA_{I\bar{z}}\ .
\loppu
Now it is easy to find the general form of
the ground state wave function. We can take the
irrelevant constant to be zero for this purpose.
The ground state wave function then satisfies
\alku
  H\psi = \Sigma_I D_I \bar{D}_I \psi = 0\ .
\loppu
Rewriting the wavefunction in the form
$ \psi(\{ z_I,\bar{z}_I\} ) = f(\{ z_I\} )
e^{ig(\{ z_I,\bar{z}_I\} )}$ where the holomorphic
part is written explicitly,
we find that the exponent $g$ must be
\alku
  g = i\frac{\pi}{\tau_y} K_{IJ} y_I y_J\ .
\loppu
Thus the ground state wavefunction is of the general form
\alku
  \psi = f(\{ z_I\}) e^{-\frac{\pi}{\tau_y} K_{IJ} y_I y_J} \ .
\loppu

\medskip

Let us now turn to examine the symmetry
properties of the Hamiltonian. We expect it to
be symmetric under translations on a torus
spanned by $2N$ complex vectors
\begin{eqnarray*}
 (z_1,\ldots ,z_N) &=& (1,0,\ldots ,0),\ (0,1,0,\ldots ,0),
\ldots ,\ (0,0,\ldots ,0,1), \\
                   & & (\tau ,0,\ldots ,0),\ (0,\tau ,0,\ldots ,0),
\ldots ,\ (0,0,\ldots ,0,\tau ) \ .
\end{eqnarray*}

First we would like to find the magnetic
translation generators. The Hamiltonian is
\alku
   H = \frac{1}{2m_0} \Sigma_I
(\vec{p}_I \cdot \vec{p}_I -
2\vec{p}_I \cdot \vec{A}_I + \vec{A}_I \cdot \vec{A}_I )\  .
\loppu
Define now
\alku \left\{ \begin{array}{l}
 \pi_{Ix} \equiv p_{Ix} - A_{Ix} =
p_{Ix} + \frac{2\pi }{\tau_y} K_{IJ} y_J \\
 \pi_{Iy} \equiv p_{Iy} - A_{Iy} =
p_{Iy}\ \end{array} \right. \ . \loppu
Then a straightforward calculation shows that
\alku
 [\pi_{Ix} - \frac{2\pi}{\tau_y} K_{IJ} y_J,\ H] = 0 \ \ ,\ \
 [\pi_{Iy} + \frac{2\pi}{\tau_y} K_{IJ} x_J,\ H] = 0 .
\loppu
Thus we can construct the following $2N$
translation generators
\alku \left\{ \begin{array}{l}
  t_{I1} =
\exp \ [ i(K^{-1})_{IJ} (\pi_{Jx} -
\frac{2\pi}{\tau_y} K_{JL} y_L) ] =
\exp \ [ i(K^{-1})_{IJ} \pi_{Jx} - i\frac{2\pi}{\tau_y} y_I ]  \\
  t_{I2} =\exp \ [ i\tau_x((K^{-1})_{IJ} (\pi_{Jx} -
\frac{2\pi}{\tau_y} K_{JL} y_L))
                  + i\tau_y((K^{-1})_{IJ} (\pi_{Jy} +
\frac{2\pi}{\tau_y} K_{JL} x_L)) ]
           \end{array} \right. \loppu
where $I = 1,\ldots ,N$. These all commute
with the Hamiltonian
\alku
 [t_{I1},H] = [t_{I2},H] = 0
\loppu
and they satisfy the following ``generalized Heisenberg algebra'':
\alku  [\ t_{I1}\ ,\ t_{J1}\ ] = 0 = [\ t_{I2}\ ,\ t_{J2}\ ] \loppu
for all $I,J = 1,\ldots ,N$ and
\alku t_{I1}t_{J2} = \exp (2\pi i (K^{-1})_{IJ}) t_{J2}t_{I1}\ . \loppu
Notice that we have arrived at the same algebra structure that
has been found in the litterature of topological Chern-Simons
theories \cite{2}.\footnote{See also \cite{13} where a similar
result has also been found for Wilson loops.}

We can also define the following translation
generators that commute with the Hamiltonian
\alku T_i = \prod^N_{I=1} t_{Ii}\ , i=1,2
\loppu
that we find to  satisfy the algebra
\begin{eqnarray}
   [ T_i,T_i] = 0\ , \ i=1,2 \nonumber \\
   T_1T_2 = e^{i2\pi \nu} T_2T_1 \ ,
\end{eqnarray}
where $\nu $ is the filling fraction.
We study these translation generators and their
representations in terms of the ground states in section 6.

\newpage

{\large {\bf Figure captions}}

\bigskip

{\bf Fig. 1a.} The action of the translation generators $t_{I2}$
in the basis I for the
$K = \left( \begin{array}{cc} 1 & 3 \\ 3 & 1 \end{array} \right)$
states. Solid line arrow = $t_{12}$-action,
dashed line arrow = $t_{22}$-action.

\medskip

{\bf Fig. 1b.} The action of the $T_2$-generator in the basis I.

\medskip

{\bf Fig. 1c.} Block diagonalization of the $T_2$-representation.

\medskip

{\bf Fig. 1d.} The $T_1,T^2_2$ quantum numbers for
the eight states of Fig. 1c. Numbering:
$$
\begin{array}{l}
   1. = \rtw [(4,4)-(0,0)] \ \ \ 2. = \rtw [(4,4)+(0,0)]  \\
   3. = \rtw [(7,3)-(3,7)] \ \ \ 4. = \rtw [(7,3)+(3,7)]  \\
   5. = \rtw [(2,2)-(6,6)] \ \ \ 6. = \rtw [(2,2)+(6,6)]  \\
   7. = \rtw [(5,1)-(1,5)] \ \ \ 8. = \rtw [(5,1)+(1,5)]
\end{array}
$$

\medskip

{\bf Fig. 1e.} The $T^2_1,T^2_2$ quantum numbers
for the four blocks of Fig. 1c.

\medskip

{\bf Fig. 2a.} The action of the translation generators $t_{I2}$
in the basis I for the
$K = \left( \begin{array}{cc} 5 & 3 \\ 3 & 3 \end{array} \right)$
states. Solid line arrow = $t_{12}$-action,
dashed line arrow = $t_{22}$-action.

\medskip

{\bf Fig. 2b.} The action of the $T_2$-generator in the basis I.

\medskip

{\bf Fig. 2c.} The $T_1,T^3_2$ quantum numbers for the
six states of Fig. 2a. Numbering:
$$
\begin{array}{l}
   1. = (0,0);\ (3,3)\\
   2. = (0,2);\ (3,5)\\
   3. = (0,4);\ (3,1)
\end{array}
$$

\medskip

{\bf Fig. 2d.} The $T^3_1,T^3_2$ quantum numbers for
the two blocks of Fig. 2b.

\newpage

\bigskip

{\large {\sc Figures}}

\bigskip


\bigskip

\newpage

\begin{thebibliography}{99}

\bibitem{01} Y.W. Suen \etal ,
Phys. Rev. Lett. {\bf 68}, 1379 (1992)\newline
             J.P. Eisenstein \etal ,
Phys. Rev. Lett. {\bf 68}, 1383 (1992)

\bibitem{02} X.G. Wen and A. Zee,
Nucl. Phys. {\bf B15}, 135 (1991);
Phys. Rev. {\bf B44}, 274 (1991); {B46}, 2290 (1992);
             Phys. Rev. Lett. {\bf 69}, 953 (1992)\newline
             J. Fr\"{o}lich and A. Zee,
Nucl. Phys. {\bf B364}, 517 (1991)\newline
             A. Zee, {\em From Semionics to Topological Fluids},
ITP preprint NSF-ITP-91-129 (1991);
             {\em Long Distance Physics of Topological Fluids},
in the proceedings of ``Low Dimensional Field Theories
             and Condensed Matter Physics'', Kyoto, 1991;
Int. J. Mod. Phys. {\bf B5}, 1629 (1991)

\bibitem{03} B.I. Halperin, Helv. Phys. Acta {\bf 56}, 75 (1983)

\bibitem{04} E. Witten, Commun. Math. Phys. {\bf 121}, 351 (1989)

\bibitem{05} For a review, see for example X.G. Wen,
Int. J. Mod. Phys. {\bf B6}, 1711 (1992)

\bibitem{1} X.G. Wen, Int. J. Mod. Phys. {\bf B2}, 239 (1990);
Phys. Rev. {\bf B40}, 7387 (1989)

\bibitem{111} G.V. Dunne, R. Jackiw and C.A. Trugenberger,
Phys. Rev. {\bf D41}, 661 (1990)

\bibitem{2} see A. Polychronakos,
Ann. Phys. (N.Y.) {\bf 203}, 231 (1990);
{\em Abelian Chern-Simons Theories and Conformal Blocks},
Univ. of Florida preprint UFIFT-HEP-89-9;
Phys. Lett. {\bf B241}, 37 (1990)\newline
                M. Crescimanno, Univ. of
California at Berkeley Ph.D. Thesis, chapter V

\bibitem{3} D. Mumford: {\em Tata Lectures in Theta}
vol. I, Birkh\"auser (1983)

\bibitem{4} F.D.M. Haldane and E.H. Rezayi,
Phys. Rev. {\bf B31}, 2529 (1985)

\bibitem{5} X.G. Wen and A. Zee, Phys. Rev. {\bf B46}, 2290 (1992)

\bibitem{6} F. Wilczek, Phys. Rev. Lett. {\bf 69}, 132 (1992)

\bibitem{61} X.G. Wen and A. Zee, Phys. Rev. Lett. {\bf 69}, 1811 (1992);
{\em Tunneling in Double Layered Quantum Hall
             Systems}, SBITP preprint NSF-ITP-92-78 (1992)\newline
             M. Greiter, X. G. Wen, F. Wilczek,
Phys. Rev. Lett. {\bf 66}, 3205 (1991); Nucl. Phys. {B374}, 567 (1992);
             {\em Paired Hall States in Double Layered Electron Systems},
preprint IASSNS-HEP-92-1 (1992)\newline
             Z.F. Ezawa, A. Iwazaki, {\em Chern-Simons Gauge
Theory for Double-Layered Electron System}, TU-402 and
             Nisho-18 (1992); {\em Quantum Hall Liquid,
Josephson Effect and Hierarchy in Double-Layered Electron System},
             TU-412 and Nisho-20 (1992)\newline
             Z.F. Ezawa, A. Iwazaki and Y.S. Wu, {\em Josephson
Effect in Multi-Layered Quantum Hall Systems}, TU-412, Nisho-21
             and CTP-2168\newline
             C. Ting, Int. J. Mod. Phys. {\bf B6}, 3155 (1992);
{\em Duality in Multi-Layered Quantum Hall Systems}, National
             University of Singapore preprint NUS-HEP-92-0503 (1992)

\bibitem{7} B. Halperin, Phys. Rev. Lett. {\bf 52}, 1583 (1984)\newline
            R.B. Laughlin, Surf. Sci. {\bf 141}, 11 (1984)\newline
            S.M. Girvin, Phys. Rev. {\bf B29}, 6012 (1984)\newline
            A.H. MacDonald and D.B. Murray, {\it ibid.},
{\bf 32}, 2707 (1985); A.H. MacDonald \etal ,
            {\it ibid.} {\bf 31}, 5529 (1985)\newline
            J.K. Jain, Phys. Rev. Lett. {\bf 63}, 199 (1989);
Phys. Rev. {\bf B40}, 8079 (1989); {\bf B41}, 7653 (1990)\newline
            B. Blok and X.G. Wen, Phys. Rev. {\bf B43}, 8337 (1991)

\bibitem{8} N. Read, Phys. Rev. Lett. {\bf 65}, 1502 (1990)

\bibitem{81} F.D.M. Haldane, Phys. Rev. Lett. {\bf 51}, 605 (1983)

\bibitem{9} R. Willet \etal , Phys. Rev. Lett. {\bf 59}, 1776 (1987)

\bibitem{91} F.D.M. Haldane and E.H. Rezayi,
Phys. Rev. Lett. {\bf 60}, 956 (1988) [Erratum: {\bf 60}, 1886 (1988)]

\bibitem{92} G. Moore and N. Read, Nucl. Phys. {\bf B360 (FS)}, 362 (1991)

\bibitem{10} E.H. Rezayi, private communication

\bibitem{11} L.J. Mordell: {\em Diophantine Equations},
Academic Press (1969)

\bibitem{12} H.M. Stark: {\em An Introduction to Number Theory},
MIT Press (1978)

\bibitem{121} F. Wilczek and A. Zee, Phys. Rev. Lett. {\bf 52}, 2111 (1984)

\bibitem{13} D. Wesolowski, Y. Hosotani and C-L. Ho,
{\em Multiple Chern-Simons Fields on a Torus}, preprint UMN-TH-1128/93,
             hepth@xxx.lanl.gov 9302121

\bibitem{14} Dingping Li, {\em Hierarchical Wave Functions and Fractional
 Statistics in Fractional Quantum Hall Effect on the Torus}, preprint
 SISSA-ISAS-58-92-EP, cond-mat@babbage.sissa.it 9212033; {\em Hierarchical
 Wave Functions of Fractional Quantum Hall Effect on the Torus}, preprint
 SISSA-ISAS-2-92-EP, cond-mat@babbage.sissa.it 9212034

\end{thebibliography}
\end{document}